\documentclass[11pt,a4paper,titlepage,openbib]{article}
\usepackage{subfig}
\usepackage{amsmath}
\usepackage{amsfonts}
\usepackage{amssymb}
\usepackage[pdftex]{graphicx}

\begin{document}
%\firstpage{1}

\title{C Library for Simulated Evolution of Biological Networks}
\author{Deepak Chandran$^{1,*}$ and Herbert M. Sauro
\footnote{Department of Bioengineering, University of Washington, Box 355061, William H. Foege Building, Room N210E, Seattle, WA, USA 98195-5061}}

\maketitle

\newenvironment{mylisting}
{\begin{list}{}{\setlength{\leftmargin}{1em}\setlength{\rightmargin}{1em}}\item\scriptsize\bfseries}
{\end{list}}

\begin{abstract}
\section{Motivation:} 
Simulated evolution of biological networks can be used to generate functional networks as well as investigate hypotheses regarding natural evolution. A handful of studies have shown how simulated evolution can be used for studying the functional space spanned by biochemical networks, studying natural evolution, or designing new synthetic networks. If there was a method for easily performing such studies, it can allow the community to further experiment with simulated evolution and explore all of its uses.

\section{Result:}
We have developed a library written in the C language that performs all the basic functions needed to carry out simulated evolution of biological networks. The library comes with a generic genetic algorithm as well as genetic algorithms for specifically evolving genetic networks, protein networks, or mass-action networks. The library also comes with functions for simulating these networks. A user needs to specify a desired function. A GUI is provided for users to become oriented with all the options available in the library.

\section{Availability: } The library is free and open source under the BSD lisence and can be obtained at evolvenetworks.sourceforge.net. It can be built on all major platforms. The code can be most conveniently compiled using cross-platform make (CMake). 
\end{abstract}

\section{Introduction}

Genetic algorithms (GA) \cite{mitchell1996iga} is an optimization algorithm that is inspired by the process of natural selection, where the fittest individuals are selected through competition. A GA is an iterative algorithm that optimizes a fitness function by repeatedly mutating the current set of individuals to create a new population of inidividuals. Each individual represents a candidate solution to the fitness function. GAs have been used in a variety of works to evolve oscillators and functioal biochemical networks \cite{deckard2004pss,paladugu2006silico} and modular networks \cite{kashtan2005sem,hintze2008evolution}. They have also been used to investigate the structure and function of genetic networks \cite{franccois2007dse, krishnan2007evolution, crombach2008evolution} and signaling networks \cite{pfeiffer2005evolution, dematte2007formal, dematt2008evolving}. A similar search algorithm, simulated annealing, has also been used to designing logical synthetic networks based on user input \cite{rodrigo2007gad}. The purpose of the library presented in this work is to allow users to experiment with simulated evolution without having to program the entire algorithm. This library can be a tool for conducting studies similar to the ones mentioned above. Previous works \cite{deckard2004pss, rodrigo2007gad} have built applications that evolve networks using a table of values provided by the user. The networks are optimized so that their steady state or time-course behavior match the input table. The advantage of a library is that the fitness function is not limited to such functions alone. Any function that can be programmed can be used as the fitness funciton. The tabular input is a small subset of the possible fitness functions. Further, the library can be used within other applications that want to utilize GAs to evolve networks. A graphical interface is also provided for quickly becoming oriented with the options available in the library. Since the library is written in the C programming language, it is easily portable and can be be extended to scripting languages such as Python, Perl, Ruby, and R. 

%, allowing complex simulations such as adaptive fitness functions \cite{loewe2009framework}

\section{Library design}

The network evolution library consists of a generic GA library and a series of extensions that provide support for simulation, export, and construction of three kinds of networks: bimolecular reactions, enzymatic reactions, protein interactions, and gene regulatory networks. The extension functions are fully configurable, allowing devlopers to replace any specific implementation with their own. The library also provides a function for tracking informaiton such as fitness distribution, evolutionary lineage, and network architecture during each generation. 

\subsection{Generic GA}

The network evolution library uses a generic GA as its foundation. This GA is very general because it uses a user-defined data structure to define an ``individual". However, the cost of such flexibility is that it requires the user to provide a number of functions, such as initializing an individual, freeing the memory occupied by an individual, mutating an individual, and performing cross-over between two individuals. The GA provides three selection functions: roulette wheel, tournament selection, and elitism. Users may add additional selection functions. While the benefit is that this GA algorithm can be highly customized, the price is that the user needs to do a substantial amount of programming. The extensions alleviate this load from the user, as discussed in the next section. 

A typical use of the generic GA library would be as follows:

\begin{verbatim}
GAinit( makeNetwork, 
        deleteNetwork, 
        fitnessFunc, 
        mutationFunc, 
        crossoverFunc, 
        selectionFunc);
        
GApopulation p = GArun( randomPopulation(), 
                      1000,   //population size
                      100,   //iterations
                      &callback ); //callback
\end{verbatim}

\begin{figure}[!h]
  \vspace{-5pt}
  \begin{center}
    \includegraphics[width=0.7\textwidth]{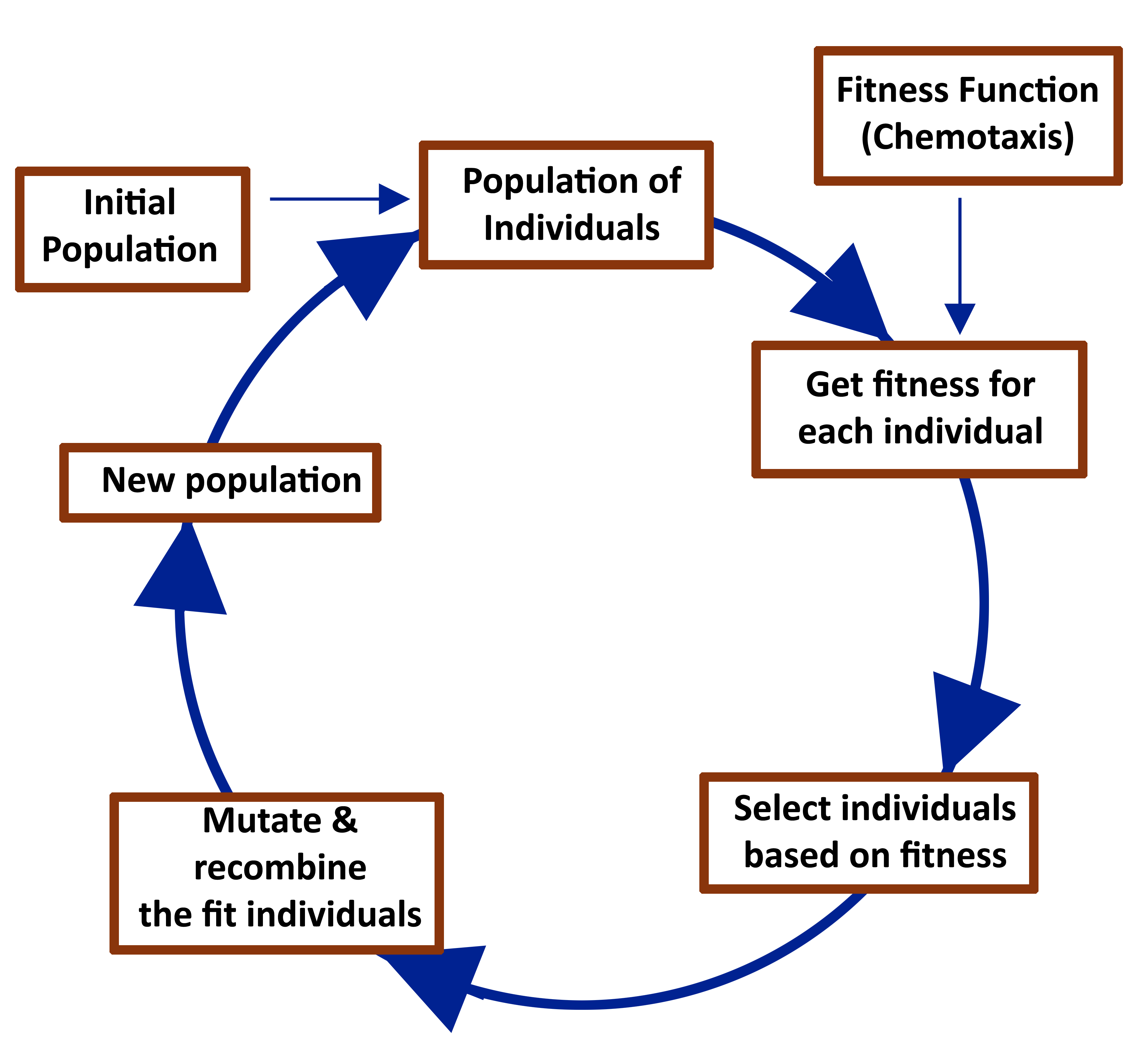}
  \end{center}
  \vspace{-5pt}
  \caption{A genetic algorithm is a simulation of natural selection. In the first step, a random set of ``individuals" is created. These individuals represent random solutions to the objective, or fitness, function. The fitness function is used to rank and select the more ``fit" individuals. The fit individuals are then mutated and recombined in order to create a new set of individuals, the next generation. The process is then repeated with this new set of individuals.}\label{fig:ga}
\end{figure}

The first function, GAinit, is used to provide the GA with the list of functions that it should use. The second function, GArun, will use these functions to perform simulated evolution and return the final population of inidividuals. The optional callback function can be used to monitor the progress of the GA as well as terminate the GA if needed. See Figure \ref{fig:ga}.

\subsection{Evolving biological networks}

The network evolution library provides data-structures and supporting functions for evolving three types of biological networks: bimolecular reaction networks, enzyme-catalyzed reaction networks, and gene regulatory networks. These three components of the library remove much of the programming load from the user. The user simply needs to define a fitness function; all the other functions required by the GA are provided. A typical use case would be as follows:

\begin{verbatim}
#include "reactionNetwork.h

//calculate fitness
double myFitness(Individual x) 
{  
    //...code...
    return f;  
}

//call-back function (optional)
int callback(int iteration,
              GApopulation p,
              int size)
{ 
   //...code...
   return 0; //return 1 to terminate
}

int main()
{
  int i;
  
  setNetworkType
  (GENE_REGULATION_NETWORK);
  
  setFitnessFunction( myFitness );
  
  //run the genetic algorithm 
  
  //1000 initial networks and 
  
  GApopulation p = evolveNetworks(
       1000,  //initial population size
       200, //successive population size
       100, //number of generations
       callBack); //call-back function
  
  printNetwork(p[0]);  //print

  for (i=0; i < 1000; ++i) //free memory
     deleteNetwork(p[i]);
}
\end{verbatim}

In the code above, all the functions except ``myFitness" and the "callback" are provided in the header file, ``reactionNetwork.h".  The \textit{setNetworkType} function allows the user to select either gene regulatory network, mass action network, enzyme networks, or protein interaction network. Alternatively, the user may also use a combination of the three network types by setting a probability of each, using the \textit{setNetworkTypeProbability} function. For example, the following code will construct a population consisting of different networks:

\begin{verbatim} 
    setNetworkTypeProbability(
    	MASS_ACTION_NETWORK, 0.2);
    setNetworkTypeProbability(
    	MASS_ACTION_NETWORK, 0.4);
    setNetworkTypeProbability(
    	PROTEIN_INTERACTION_NETWORK, 0.2);
    setNetworkTypeProbability(
    	GENE_REGULATION_NETWORK, 0.2);    
\end{verbatim}

\subsubsection{Simulating the networks}

Since the three networks provided in the library represent reaction networks, the library also provides functions for simulating a given network, both stochastically and deterministically. Stochastic simulations are performed by a custom Gillespie algorithm \cite{gillespie1977es}, and the deterministic simulation is performed by the Sundials library \cite{hindmarsh2005ssn}. The simulation functions are included with the library. Functions for computing the reaction rates and stoichiometry matrix for each network type are defined in the library. These functions are used for the simulations. The user can use custom reaction rate laws by overriding these functions using \textit{setRatesFunction} and \textit{setStoichiometyFunction}.

\subsubsection{Configuring the networks}

Just as the reaction rate equations for each network type can be modified using \textit{setRatesFunction}, other functions, such as the selection, mutation, and crossover, can also be replaced with custom functions. In addition, the randomly generated networks can be configured. For example, the function \textit{setParametersForGeneRegulationNetwork} can be used to configure various properties of gene regulation networks, such as the average number of regulations per gene or the average degradation rate. Analogous functions exist for configuring aspects of bimolecular networks and protein interaction networks. The average network size can also be configured using \textit{setNetworkSize}.

\subsubsection{Exporting the networks}

Each of the three network types included in the library can be exported to Jarnac or Antimony format \cite{sauro2003ng, smith2009antimony} by using the \textit{printNetwork} function. These file formats are text-based languages for representing reaction networks using reactions and rate expressions, and they can be converted to the Systems Biology Markup Language \cite{hucka2003sb} using the Systems Biology Workbench \cite{sauro2003ng}.

\subsection{Available network types}

The three network types that are defined in the network evolution library are described in the next three sections. 

\subsubsection{Bi-molecular reaction networks}

\begin{figure}[!h]
  \vspace{-10pt}
   \centering
    \subfloat[metabolic network]{\label{fig:metabolic_model}\includegraphics[width=0.3\textwidth]{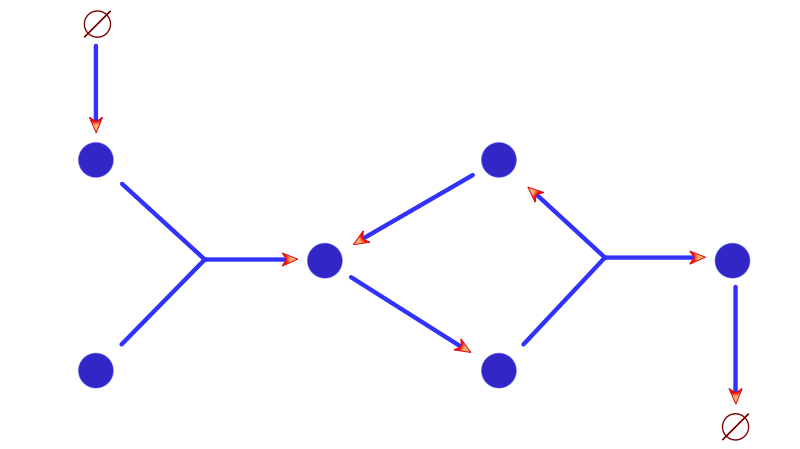}}                
    \subfloat[protein network]{\label{fig:protein_model}\includegraphics[width=0.3\textwidth]{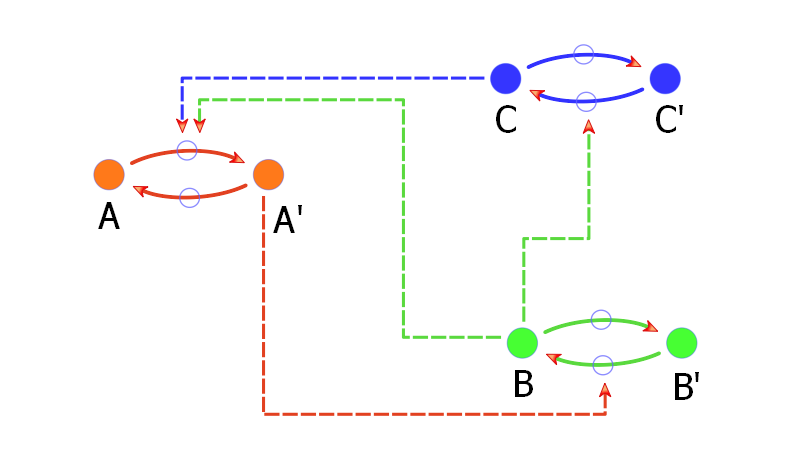}}
    \subfloat[genetic network]{\label{fig:genetic_model}\includegraphics[width=0.3\textwidth]{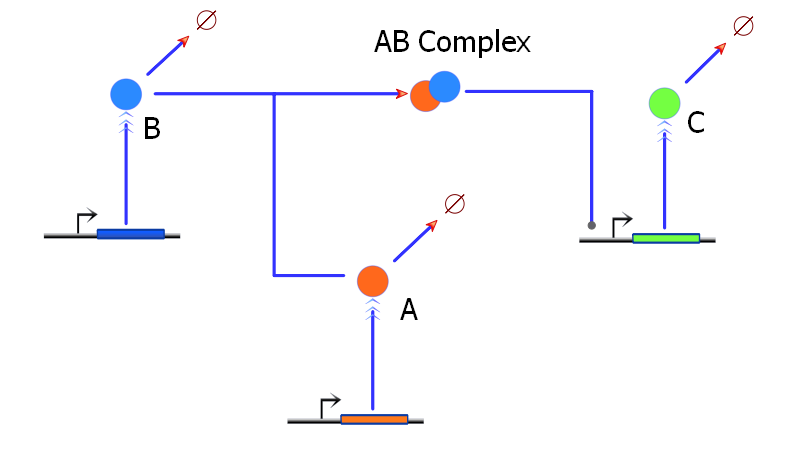}}
  \vspace{-10pt}
  \caption{ The network evolution library consists of three types of networks, each modeled using different kinetics: (a) Metabolic network using mass-action kinetics (b) Protein network using Michaelis-Menten kinetics (c) Gene regulatory network using fractional saturation models.}\label{fig:types_of_networks}
\end{figure}

The library defines a biochemical network as a set of unimolecular or bimolecular reactions obeying mass-action kinetics. In other words, each reaction is of the form $A + B -> C + D$, where $A$ and $B$ are the reactants and $C$ and $D$ are the products. The number of reactants and products can be at most two. Zero products or reactants (not both) are allowed, because this represents some flux to or from an external source. See Figure \ref{fig:types_of_networks}(a).

\subsubsection{Enzymatic reaction networks}

The library extends the bi-molecular reaction networks by including a new rate expression for reactions with a single substrate and a single product. For reactions with single a substrate and product, the reversible Hill equation is used for the rate law \cite{hofmeyr1997reversible}. The enzymatic reaction network library illustrates the extensibility of the entire library, because it is a network structure that was built on an existing network structure, the bi-molecular reactions network. The additional information required for defining the reversible Hill equation was included in the structure. Note that this type of network cannot be simulated using the Gillespie algorithm \cite{gillespie1977es}, since the rates can be negative. 

\subsubsection{Networks with protein state changes}

The library defines a protein interaction network as a set of variables, i.e. proteins, that can switch between active and inactive states. The physical interpretation of these states can be conformational changes or covalent modifications. The transition to and from active state is controlled by other proteins in the network. The number of enzymes controlling each transition can be none or many. See Figure \ref{fig:types_of_networks}(b). 

\subsubsection{Gene regulatory network}

The library defines a gene regulatory network as a set of variables, i.e. transcription factors, that have a production rate and degradation rate. Each gene product is controlled by a number of complexes. Each complex can be composed of one or many gene products. See Figure \ref{fig:types_of_networks}(c). Each gene product has a corresponding production rate and degradation rate. The degradation rate is a linear function, and the production rate uses the fractional saturation model, which assumes that the binding and unbinding of transcription factors to the operator sites are at equilibrium \cite{shea1985oc}. As mentioned earlier, these default rate laws can be overridden by the user using the \textit{setRatesFunction} function.

\begin{figure}[!h]
  \vspace{-10pt}
   \centering
    \subfloat[evolved oscillator]{\label{fig:evolvedoscil}\includegraphics[width=0.5\textwidth]{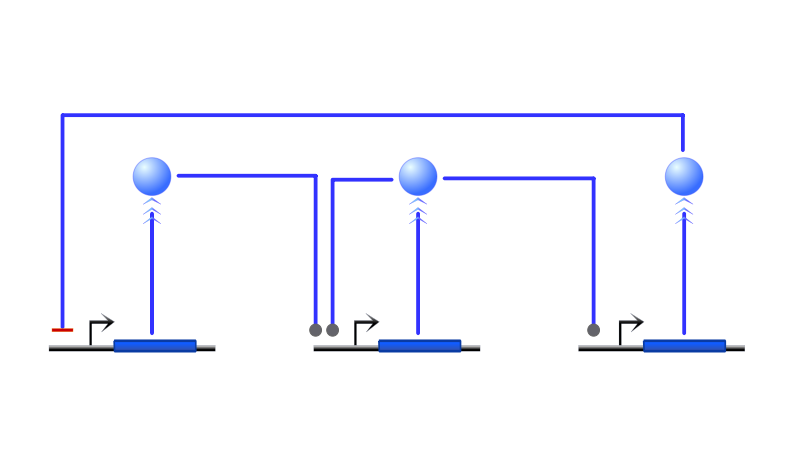}}                
    \subfloat[evolved XOR gate]{\label{fig:evolvedxor}\includegraphics[width=0.3\textwidth]{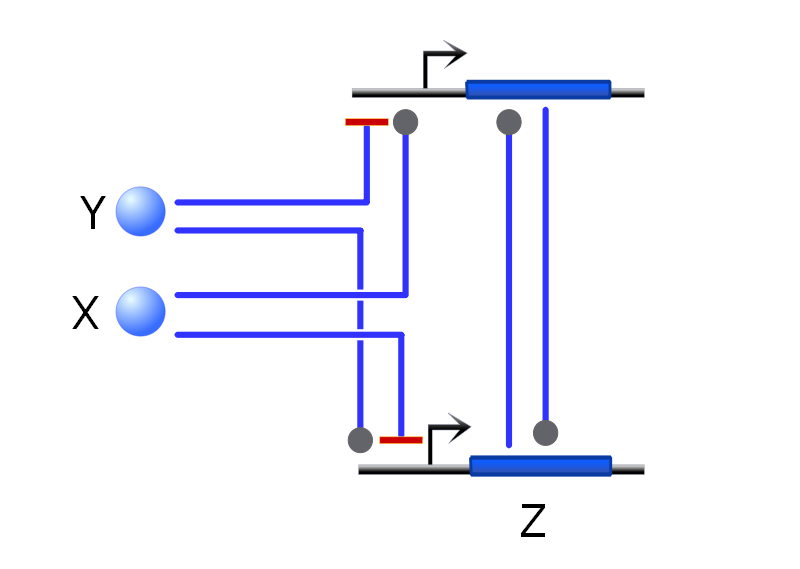}}
  \vspace{-10pt}
  \caption{ Shown here are two networks that were evolved using the network evolution C library; specifically, the gene regulatory network evolution portion of the library was used.  (a) An evolved oscillator network that relies on a negative feedback composed of two positive regulators (grey circle) and one negative regulator (red bar). (b) An evolved XOR logic gate where X and Y are the inputs and Z is the output. The steady state behavior of the network is such that Z is only upregulated when either X or Y is high, but not both. The network relies on a positive feedback loop that is triggered as long as one of the genes is upregulated. When both X and Y are high, both genes are downregulated, preventing the positive feedback from taking effect.}\label{fig:evolvedgrns}
\end{figure}

%As an example, suppose X is the target transcription factor that is controlled by two other transcription factors: A is an activator, and the dimer formed from B and C is an inhibitor. Then the production rate of X will be written as: $k0 * k1 * A / ( 1 + k1 * A + k2 * B * C )$. In this formula, $k0$ represent the maximum production rate. The dimer of $B$ and $C$ is represented as the product $B*C$ in the demoninator because of the equilibrium assumption, $BC = k2 * B * C$. 

%The network is defined by an array of arrays, which represent all the complexes. Each complex has an associated target gene and rate constants. Mutations are performed similar to the previous models, i.e. by altering one of the genes, target genes, compelexes, rate constants, or by adding or removing new genes or reactions. Recombination is also performed in a similar manner as the earlier models by taking random regions of two networks and merging them to form a new network.

\section{Testing the network evolution library}

Several test were performed in order to test the library. These include evolving logic gates using genetic networks \cite{rodrigo2007gad} and oscillating networks using all three different types of networks \cite{deckard2004pss}. See Figure \ref{fig:evolvedgrns}. The network evolution library was successfully able to evolve logical gates and oscillating networks. In general, one generation with 500 individuals takes approximately 1 second on an AMD Opteron 1.8Gz computer. Between 30 and 60 generations were required to evolve oscillating networks, and between 20 and 40 to evolve logic gates. 

Two relatively more complex functions were also tested: evolving networks to reduce the coefficient of variation and evolving networks to exhibit bistability \cite{chandran2009optimization}. The C code for evolving noise-reducing networks is provided in supplementary materials.

\section{Flexibility, extensibility, and limitations}
Different components of the library, such as the function for ramdomly selecting the fit individuals or the function for mutating individuals, can be overridden by using ``set" functions. Similarly, a user can change the rate laws for the three network types. The limitation is that some of these changes require the programmer to understand the underlying data structure used in the library. 

It is also possible to define an entirely new network type other than the three provided by the library. Although this would require a substantial amount of programming, the advantage is that the programmer can integrate the new network into the library. For example, suppose that the programmer designs a new network type called \textit{My\_Network\_Type}. In order to fully define this network type, the programmer would provide functions for constructing, removing, mutating and performing crossover on \textit{My\_Network\_Type}, and simulating and printing a \textit{My\_Network\_Type}. Once these functions are written, the programmer may add the functions into the existing library using an index value, say \textit{MY\_NETWORK}. A new user can use this new network type by using \textit{setNetworkType(MY\_NETWORK)}.

\section{Conclusions}

This work presents a C library for evolving three different types of reaction networks using genetic algorithms, which include networks composed of bimolecular reactions, protein interaction networks, and gene regulatory networks. The library is capable of evolving oscillators and logic gates as well as more complex functions. The ability to program the fitness function maximizes the types of simulated evolution experiments one can perform. Since the library is written in the C programming language, it is relatively simple to create an interface for scripting languages such as Python, Perl, Ruby, or R, and even higher level languages such as Java using SWIG. 

\section{Acknowledgements}
This work was funded by the National Science Foundation (ID 0527023- FIBR).

\pagebreak 
\section{Literature Cited}
\renewcommand{\refname}{}
\bibliographystyle{plain}
\bibliography{networkEvolutionLib}
\pagebreak 

\label{suppl}
\section*{Supplementary information}

The following code can be used to evolve networks that minimize the noise in a network. The fitness function is defined as the inverse of the coefficient of variation. The best network will have the lowest coefficient of variation.

\begin{verbatim}

#include "reactionNetwork.h"

/* fitness = 
  coefficient of variation (CV) */
double fitness(void * p);

/* print the generation number 
and fitness of best network during each iteration */

int callback(int iter,GApopulation pop,int popSz);

/*Main*/
int main()
{	
	setFitnessFunction( &fitness );  
	setNetworkType( GENE_REGULATION_NETWORK );
	setInitialNetworkSize(4,3); 
	
	/*evolve using 1000 initial networks, 
	 500 neworks during each successive generation,
	 for 20 generations */
	GApopulation pop = evolveNetworks(200,
	                                  100,
	                                  20,
	                                  &callback);  
	
	 //get the best network
	ReactionNetwork * net = (ReactionNetwork*)pop[0];
	
	printNetwork(net); //print the best network
	
	/*simulate and write the result*/
	
	int i;
	int r = getNumReactions(net);
	int n = getNumSpecies(net);
	
	//initial values
	double * iv = malloc( n * sizeof(double));
	for (i = 0; i < n; ++i) iv[i] = 0.0;

    //stochastic simulation
	int sz;
	double * y = 
	    simulateNetworkStochastically
	         (net,iv,500,&sz);
	         
	free(iv);
	
	//write table to file
	writeToFile("dat.txt",y,sz,n+1); 
	
	free(y);
	
	/*free all the networks*/
	
	for (i=0; i < 100; ++i)
		deleteNetwork(pop[i]);
		
	return 0;
}

/* fitness = 
   coefficient of variation (CV) */
double fitness(void * p)
{
	int i;
	
	//the network to test
	ReactionNetwork * 
	   net = (ReactionNetwork*)p;
	
	int n = getNumSpecies(net);
	int r = getNumReactions(net);
	
	//initial concentrations
	double * iv = malloc
	    ( n * sizeof(double));
	for (i = 0; i < n; ++i) 
	    iv[i] = 0.0;

	double time = 500.0;
	
	//stochastic simulation
	int sz;
	double * y = 
	   simulateNetworkStochastically
	        (net,iv,time,&sz);
	free(iv);

    //compute the variance
	double f = 0, sd, dt;
	if (y != 0)         
	{
		double mXY = 0,mX = 0,
               mY = 0, mX2 = 0, 
		       mY2 = 0;
		for (i = 0; i < (sz-1); ++i)
		{
			dt =   getValue(y,n+1,i+1,0) 
			     - getValue(y,n+1,i,0);
			mX +=  getValue(y,n+1,i,1) 
			       * dt;
			mX2 += getValue(y,n+1,i,1)
			       * getValue(y,n+1,i,1)
			       * dt;
		}
		mX /= time;
		mX2 /= time;
		
		//standard deviation
		sd = sqrt(mX2 - mX*mX);  
		
		if (sd <= 0 || 
		    mX <= 0 || mX > 5.0) 
			f = 0.0;
		else
			f = mX / sd;   //CV
		
		free(y);
	}
	
	//disallow large networks
	if(getNumSpecies(net) > 5) 
		f = 0.0;
	
	return (f);  //CV
}

/* print the generation number and fitness of best network during each iteration */
int callback(int iter,GApopulation pop,int popSz)
{
	int i,j;
	double f = fitness(pop[0]);
	ReactionNetwork * net = (ReactionNetwork*)(pop[0]);
	
	printf("%i\t%i\t%lf\n",iter,getNumSpecies(net),f);
	if (iter > 50 && f < 0.5)
	{
		for (i=1; i < popSz; ++i)
		{
			for (j=0; j < 10; ++j)
				pop[i] = mutateNetwork(pop[i]);
		}
	}
	return 0;
}

\end{verbatim}

\end{document}